\global\def\draftcontrol{0}

   \def\versionno{ holomorphic anomaly }
\catcode`\@=11

\expandafter\ifx\csname draftcontrol\endcsname\relax\global\def\draftcontrol{1} 
\fi 

{\count255=\time\divide\count255 by 60 
\xdef\hourmin{\number\count255} 
\multiply\count255 by-60\advance\count255 by\time 
\xdef\hourmin{\hourmin:\ifnum\count255<10 0\fi\the\count255}} 
\def\draftdate{\number\month/\number\day/\number\year\ \ \ \hourmin } 

\newcommand\makepapertitle{\par

  \begingroup 
    \renewcommand\thefootnote{\@fnsymbol\c@footnote}% 
    \def\@makefnmark{\rlap{\@textsuperscript{\normalfont\@thefnmark}}}% 
    \long\def\@makefntext##1{\parindent 1em\noindent 
            \hb@xt@1.8em{% 
                \hss\@textsuperscript{\normalfont\@thefnmark}}##1}% 
     \newpage 
     \global\@topnum\z@   % Prevents figures from going at top of page. 
     \@makepapertitle 
     \thispagestyle{empty}\@thanks 
  \endgroup 
  \setcounter{footnote}{0}% 
  \global\let\thanks\relax 
  \global\let\makepapertitle\relax 
  \global\let\@makepapertitle\relax 
  \global\let\@thanks\@empty 
  \global\let\@author\@empty 
  \global\let\@date\@empty 
  \global\let\@title\@empty 
  \global\let\title\relax 
  \global\let\author\relax 
  \global\let\date\relax 
  \global\let\and\relax 
  \def\version{\let\version\@version\@gobble} 
} 
\def\@makepapertitle{% 
  \newpage 
   \ifnum\draftcontrol=1 {} 
   \version\versionno 
   \vskip 7em% 
   \else 
   \hfill\hbox to 3cm {\parbox{4cm}{\@pubnum}\hss}% 
   \vskip 7em% 
   \fi 
   \begin{center}% 
   \let \footnote \thanks 
      {\hskip -0\textwidth \hbox to 1\textwidth% 
        {\centerline{\Large\bf{\noindent\@title}}}}% 
     \vskip 1.5em% 
     {\normalsize%\large 
       \lineskip 1.5em% 
       \begin{tabular}[t]{c}% 
         \@author 
       \end{tabular}\par}% 
     \vskip 5.5em% 
     {\@bstract}% 
     \end{center}% 
     \vfill
     \@date%
     \vskip 1.5em%
   \par 
} 

\gdef\@pubnum{} 
\def\pubnum#1{% 
  \gdef\@pubnum{#1}} 

\gdef\@bstract{} 
\def\Abstract#1{% 
  \gdef\@bstract{% 
   \parbox{\textwidth-3pc}{% 
   \centerline{\bf Abstract}\penalty1000 
   \renewcommand\baselinestretch{1.0} 
   \noindent
   {#1}}} 
} 

\gdef\@email{}
\def\email#1{%
   \gdef\@email{%
   Email: {\tt #1}}
}

\def\ps@paper{\let\@mkboth\@gobbletwo% 
     \ifnum\draftcontrol=1 
        \def\@oddfoot{\hbox to \textwidth{\tiny \versionno \hfil\tiny\draftdate}% 
        \hskip -\textwidth \hbox to \textwidth{\hfil\rm\thepage\hfil}}% 
     \else\def\@oddfoot{\hbox to \textwidth{\hfil\rm\thepage\hfil}} 
     \fi 
     \let\@evenfoot\@oddfoot 
} 
\def\body{\clearpage 
          \pagestyle{paper} 
        } 
\newenvironment{acknowledgments}{% 
\vskip 3.25ex 
\noindent {\bf Acknowledgments} 
} 

\def\@version#1{\ifnum\draftcontrol=1 
\typeout{}\typeout{#1}\typeout{} 
\vskip3mm\centerline{\hbox{\fbox{\normalsize{\tt DRAFT -- #1 -- } 
                   {\draftdate}}}}\vskip3mm 
\fi} 
\let\version\@version 
\long\def\eqlabel#1{\ifnum\draftcontrol=1 
                    \tag@false  % there are some problems with multline without this 
                    \tag*{(\theequation) \hbox to -0.2cm{\hspace{0cm}\small{#1}\hss}} 
                    \refstepcounter{equation}  
                    \edef\@currentlabel{\theequation} 
                    \ltx@label{#1}          % use old LaTeX \label instead of new definition 
                    \else 
                    \label{#1} 
                    \fi 
                    } 
\let\st@bibitem\@bibitem 
\let\st@lbibitem\@lbibitem 
\ifnum\draftcontrol=1 
  \def\@bibitem#1{% 
    \st@bibitem{#1}\a@@label{#1}\ignorespaces} 
  \def\@lbibitem[#1]#2{% 
    \st@lbibitem[#1]{#2}\a@@label{#2}\ignorespaces} 
  \def\a@@label#1{% 
    \gdef\a@lab{\smash{\normalfont\small#1}} 
    \ifvmode 
      \if@inlabel 
        \global\setbox\@labels\hbox{% 
          \llap{\a@lab\let\a@lab\relax 
                \kern\@totalleftmargin\kern\marginparsep}% 
          \box\@labels}% 
      \fi 
    \fi} 
\fi 
\documentclass[letterpaper,12pt]{article} 
\usepackage{amsmath,bm,amsfonts,amssymb,array,calc,amsthm,rotating}
\usepackage{epsfig} 
\usepackage{color}
\usepackage[colorlinks=false]{hyperref}
\usepackage[margin=1in]{geometry}
\usepackage{tabularx}
\renewcommand\baselinestretch{1.25} 
\definecolor{refcol}{rgb}{0.2,0.2,0.8}
\definecolor{eqcol}{rgb}{.6,0,0}
\definecolor{purple}{cmyk}{0,1,0,0}

\gdef\@citecolor{refcol}
\gdef\@linkcolor{eqcol}
\def\colorlinkspurple{\gdef\@urlcolor{purple}}
\def\colorlinksblue{\gdef\@urlcolor{blue}}
\def\colorlinksred{\gdef\@urlcolor{red}}

\newcommand{\pc}[1]{\parbox{0pt}{#1}}

\catcode`\@=12 

\begin{document}

%%% 
%%%%%% text starts here 
%%%%%%%%% 

\title{
\parbox{\textwidth}{\begin{center}New Anomalies in Topological String Theory\end{center}}}

\pubnum{%
CALT-68-2672\\IPMU 08-0020}
\date{April 2008}

\author{
Paul L. H. Cook,$^1$ Hirosi Ooguri,$^{1,2}$ and Jie Yang$^1$ \\[0.5cm]
\it $^1$~California Institute of Technology, Pasadena, CA 91125, USA
\\[0.1cm] \it $^2$ Institute for the Physics and Mathematics of the Universe
\\ \it ~~University of Tokyo, Kashiwa, Chiba 277-8586, Japan
}

\Abstract{We show that the topological string partition function with 
D-branes on a compact Calabi-Yau manifold has new
anomalies that spoil the recursive structure of the holomorphic 
anomaly equation and introduce dependence on wrong moduli 
(such as complex structure moduli in the A-model), 
unless the disk one-point functions vanish.
 This provides a microscopic explanation for the recent result of 
Walcher in arXiv:0712.2775 on counting of BPS states in M-theory using the topological 
string partition function. The relevance of vanishing disk one-point functions to
large $N$ duality for compact Calabi-Yau manifolds is noted. 
\\
\\{\em PACS}: 11.25.Mj }

\makepapertitle

\body

\version\versionno

\vskip 1em

\newpage

\section{Introduction}

The topological string holomorphic anomaly equation gives a recursion relation
for the partition function $F_g$ 
with respect to the genus $g$ of the string worldsheet \cite{Bershadsky:1993cx}.
The equation has proven to be useful in evaluating topological 
string amplitudes. In fact, for {\it compact} Calabi-Yau
manifolds, it is the only known method for
computing these amplitudes systematically for higher $g$. 
This method has seen remarkable progress in 
recent years. The Feynman diagram method developed in \cite{Bershadsky:1993cx} has been made 
more efficient by \cite{Yamaguchi:2004bt}. This, combined with knowledge on the behavior of $F_g$ 
at boundaries of the Calabi-Yau moduli space, has made it possible to integrate 
the holomorphic anomaly equation to very high values of $g$ \cite{Huang:2006hq}.

Recently, Walcher generalized the holomorphic anomaly equation
to the case of topological string theory in the presence of D-branes \cite{Walcher:2007tp}.
Attempts to derive such an equation had been made before, 
for example in \cite{Bershadsky:1993cx}. The new ingredients in \cite{Walcher:2007tp} are
two assumptions: that open string moduli do not contribute
to factorisations in open string channels and that disk 
one-point functions vanish. A Feynman diagram method for integrating
the holomorphic anomaly equation in the presence of D-branes has subsequently
been proven \cite{Cook:2007dj} and enhanced \cite{Alim:2007qj,Konishi:2007qx}, as well
as considered in the context of background independence \cite{Neitzke:2007yw}. Furthermore,
initial attempts have been made to understand the situation where open string moduli may
contribute \cite{Bonelli:2007gv}.

In this paper, we will focus on the assumption of vanishing disk
one-point functions. We find that, for a {\it compact} Calabi-Yau
manifold, disk one-point functions generate new terms in
the holomorphic anomaly equation and spoil its recursive
structure. Moreover, with non-zero disk one-point functions, 
the string amplitudes can develop dependence on ``wrong'' moduli, that is complex
structure moduli in the A-model and K\"ahler moduli in the B-model.

That disk one-point functions themselves depend on wrong moduli
has been known for a long time. In \cite{Ooguri:1996ck}, it was shown that 
D-branes in the A-model are associated to Lagrangian 3-cycles and that
their disk one-point functions depend on B-model moduli. 
Conversely, D-branes in the B-model are associated to holomorphic
even-cycles and their disk one-point functions depend on A-model moduli.
One might then imagine that disk one-point functions could introduce wrong moduli dependence
into higher genus partition functions, and indeed we will find this effect explicitly,
as new anomalies in compact Calabi-Yau manifolds. 

The cancellation of overall D-brane charge provides a means to remove the contribution
of the new anomalies. Indeed, such a cancellation appears to be required 
for the successful counting of the number of BPS states in M-theory using the topological
string partition function. In \cite{Gopakumar:1998ki}, it was conjectured
that the partition function of the closed topological string
can be interpreted as counting BPS states in M-theory
compactified to five dimensions on a Calabi-Yau manifold. 
This conjecture was extended to cases
with D-branes in \cite{Ooguri:1999bv,Ooguri:2002gx}. Recently, Walcher \cite{Walcher:2007qp}
applied the formulae of \cite{Ooguri:1999bv} to examples of {\it compact}
Calabi-Yau manifolds and found that the integrality of
BPS state counting can be assured only when the topological charges
of the D-branes were cancelled by introducing orientifold planes \cite{Walcher:2007qp}, 
such that the disk one-point functions vanish.
Our result gives a microscopic explanation of this observation.

Furthermore, the absence of the new anomalies appears to be a prerequisite for large $N$ duality between
open and closed topological string theories. Specifically, duality implies that topological
string amplitudes in both theories should obey the same equations, notably the holomorphic
anomaly equation, and should not depend on the wrong moduli. In \cite{Cook:2007dj}, it was pointed out 
that the holomorphic anomaly equation for the open string derived in \cite{Walcher:2007tp} under the assumption
of vanishing disk one-point functions is similar to that for the closed string, after appropriate shifts of
closed string moduli by amounts proportional to the 't Hooft coupling. The compatibility of the 
holomorphic anomaly equation and large $N$ duality are further investigated in \cite{AOV}. 
Conversely, the presence of the new anomalies is correlated with
the breakdown of large $N$ duality. For compact Calabi-Yau manifolds, 
the conifold transition requires homology relations among vanishing cycles \cite{Greene:1995hu, Greene:1996dh}.
For example, if a single 3-cycle of non-trivial homology shrinks and the singularity is blown up, 
the resulting manifold cannot be K\"ahler. Thus, the presence of D-branes with 
non-trivial topological charge implies a topological string theory without closed string dual ---
and simultaneously the disk one-point functions do not vanish, and so the new anomalies are present.

In section 2, we analyse the anomalous dependence of the partition function
on both K\"ahler and complex structure moduli, and identify 
the terms expressing the new anomalies. In section 3, we discuss 
further the implications of these new anomalies. 

\section{Anomalous worldsheet degenerations}

In this section we consider the dependence of the genus $g$ and boundary number $h$ amplitude $F_{g,h}$ on both the anti-holomorphic moduli $\bar t^{\bar i}$ and the ``wrong'' moduli $y^{a}$ and $\bar y^{\bar a}$. Our results are independent of choosing the A- or B-model, so wrong moduli are complex structure moduli for the A-model and K\"ahler moduli for the B-model.

Our analysis below completes the derivation of \cite{Walcher:2007tp}, namely that under the assumption that open string moduli do not contribute to open string factorisations,
\begin{align}
  \label{eq:tbar_anomaly}
  \frac{\partial}{\partial \bar t^{\bar i}} F_{g,h} &= \frac12 C_{\bar i \bar j \bar k} e^{2K} G^{\bar j j} G^{\bar k k} \left( \sum_{r=0}^{g} \sum_{s=0}^{h} D_j F_{r,s} D_k F_{g-r,h-s} + D_jD_k F_{g-1,h} \right) - \Delta_{\bar i}^j D_j F_{g,h-1},\\
\label{eq:y_anomaly}
  \frac{\partial}{\partial y^a}  F_{g,h} &= \frac{\partial}{\partial \bar y^{\bar a}}  F_{g,h} = 0,
\end{align}
if and only if
\begin{equation}
  \label{eq:disk_one_point_function_vanishing}
   \overline C_{\bar a} = \langle \bar \omega_{\bar a} | B \rangle = 0.
\end{equation}
Here $B$ is the boundary and $\overline C_{\bar a}$ is a disk one-point function, with $\bar \omega_{\bar a}$ an $(a,a)$ chiral primary state with charges $q_{\bar a}+\bar q_{\bar a}=-3$. If the disk one-point functions do not vanish, then all three of these equations receive anomalous contributions, which cannot be written in terms of lower genus amplitudes. In the presence of orientifolds, $\overline C_{\bar a}$ is the sum of the disk and crosscap one-point functions \cite{Walcher:2007qp}.

We denote the supercurrents $G^\pm$ and $\overline G^\pm$, with conventions such that for both models the BRST operator is written as,
\begin{equation}
  \label{eq:BRST}
  Q_{BRST} = \oint G^+dz + \oint \overline G^+d\bar z,
\end{equation}
where barred quantities are right-moving. The appropriate worldsheet boundary conditions for the supercurrents are,
\begin{equation}
  \label{eq:boundary_condition}
  (G^+dz + \overline G^+d\bar z)|_{\partial\Sigma} = 0, \qquad \hbox{and} \qquad (G^-\chi^z dz + \overline G^-\bar\chi^{\bar z} d\bar z)|_{\partial\Sigma}=0,
\end{equation}
where $\chi$ is a holomorphic vector along the boundary direction.

To derive the extended holomorphic anomaly equation we follow the approach of \cite{Bershadsky:1993cx}, with the addition of some important details. Taking the $\bar t^{\bar i }$ derivative of $F_{g,h}$ is equivalent to inserting the operator,
\begin{equation}
  \label{eq:tbar_insertion}
 \int_\Sigma \{G^+, [\overline G^+, \bar\phi_{\bar i}]\},
\end{equation}
with integral over the worldsheet $\Sigma$. $\bar\phi_{\bar i}$ is a state in the $(a,a)$ chiral ring with left- and right-moving $U(1)_R$ charge $(-1,-1)$. Here $[\overline G^+,\bar\phi_{\bar i}]$ means $\oint_{C_z}dw \; \overline G^+(w)\bar\phi_{\bar i}(z),$ with $C_z$ a small contour surrounding $z$. In general the integrals of $G^+$ and $\overline G^+$ do not annihilate the boundary, so to derive the holomorphic anomaly we rewrite the insertion as,
\begin{equation}
  \label{eq:changed_insertion}
-\frac12\int_{\Sigma} \{G^++\overline G^+, [G^+ - \overline G^+, \bar\phi_{\bar i}] \},
\end{equation}
to allow at least one contour to be deformed around the worldsheet.

For the other case, namely the wrong moduli, taking the $y^a$ derivative of $F_{g,h}$ is equivalent to inserting,
\begin{equation}
  \label{eq:ybary_insertions}
\int_\Sigma \{\overline G^+, [G^-, \varphi_{a}]\}+2\int_{\partial\Sigma} \,\varphi_{a},
\end{equation}
where $\varphi_a$ is a charge $(1,-1)$ marginal operator from the $(c,a)$ ring, which satisfies $[G^+,\varphi_a]\nolinebreak=\nolinebreak0$ and $ [\overline G^-, \varphi_a] = 0$. The second term is a boundary term required to resolve the so-called Warner problem \cite{Warner:1995ay}: we require the deformation to be BRST-exact, but the $Q_{BRST}$ variation of the first term alone is a boundary term, as can be seen using $\{G^+, G^-\}= 2T$ and converting $T$ to a total derivative. Since $\int_\Sigma \{G^+, [G^-, \varphi_a]\} = 2\int_{\partial \Sigma} \varphi_a$, we can re-write (\ref{eq:ybary_insertions}) as,
\begin{equation}
\int_\Sigma\{G^++ \overline G^+, [G^-, \varphi_{a}]\}, \label{eq:changed_y_insertions}
\end{equation}
so that the first contour can be deformed past boundaries on the worldsheet.

Thus for both the $\bar t^{\bar i}$ and $y^a$ derivatives, the combination $(G^++\overline G^+)$ can be moved around the Riemann surface, producing terms corresponding to all possible degenerations of the Riemann surface, as listed below. For each degeneration, there remain the insertions,
\begin{equation}
  \label{eq:remaining_insertions}
  -\frac12 \int_\Sigma [G^+ - \overline G^+, \bar\phi_{\bar i}]\quad \hbox{ and }\quad \int_\Sigma [G^-, \varphi_{a}],
\end{equation}
for $\bar t^{\bar i}$ and $y^a$ dependence respectively.

The first class of degenerations are {\em closed string factorisations}, corresponding to a handle degenerating, and either splitting the Riemann surface in two, or removing a handle. The remaining modulus of the handle is represented by an integrated $(G^- - \overline G^-)$ insertion, folded with a Beltrami differential. This insertion annihilates the ground states propagating on the long tube, so for non-zero result the remaining insertion (\ref{eq:remaining_insertions}) must be on the tube. Now the absence of boundaries on the tube makes the results of \cite{Bershadsky:1993cx} directly applicable, namely: for the $\bar t^{\bar i}$ derivative we get all but the last term in (\ref{eq:tbar_anomaly}); and the $y^a$ derivative contributions vanish.

Next are {\em open string factorisations}, where a boundary expands and meets itself, removing a handle; or two boundaries collide. The degeneration produces a thin strip, with each end encircled by a $G^-$ or $\overline G^-$ folded with a Beltrami differential, associated with the position of the attachment point of the strip to the boundary. The strip can be replaced by a complete set of open string states. However, our assumption that open string moduli do not contribute removes all but charge $0$ and $ 3$ states, and these are annihilated by the $G^-$ or $\overline G^-$ integrated around the attachment point, regardless of the location of the insertion (\ref{eq:remaining_insertions}). Thus this case gives no contribution.

The interesting case is that of a {\em boundary shrinking}, or equivalently moving far from the rest of the Riemann surface. That such a degeneration is part of the boundary of moduli space can be seen by doubling the Riemann surface $\Sigma_{g,h}$ to form a closed surface $\Sigma'_{2g+h-1,0}$, with the boundaries of $\Sigma_{g,h}$ on the fixed plane of the $\mathbb{Z}_2$ involution of $\Sigma'_{2g+h-1,0}$. The pinching off of a $\Sigma'_{2g+h-1,0}$ handle which crosses the $\mathbb{Z}_2$ fixed plane is equivalent to a shrinking boundary in $\Sigma_{g,h}$.

A boundary is associated with three real moduli insertions, specifying the location of the boundary and its length $\tau$. The boundary degeneration is thus equivalent to a boundary at the end of a long tube, with the Beltrami differentials associated with the two remaining moduli localised to the attachment point of the tube to the rest of the Riemann surface. The absence of additional moduli on the tube distinguishes this class from the closed string factorisation class above, and furthermore allows the remaining insertion (\ref{eq:remaining_insertions}) to be anywhere on the worldsheet.

Firstly, (\ref{eq:remaining_insertions}) may nevertheless be on the tube. The degeneration $\tau\rightarrow \infty$ projects the intermediate states on both sides of the insertion to ground states, since excited states decay as $e^{-h\tau}$ where $h>0$ is the total (left+right) conformal weight. Now, however, $G^\pm$ and $\overline G^\pm$ annihilate the ground states, so this case is zero.

\begin{figure}
\begin{minipage}[t]{0.4\textwidth}
  \centering
  \begin{picture}(0,0)%
    \includegraphics{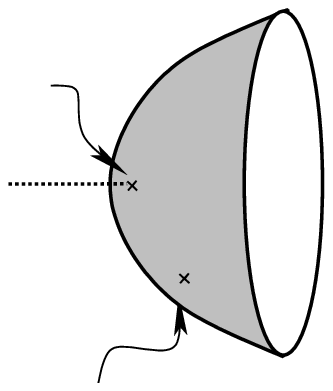}%
  \end{picture}%
  \setlength{\unitlength}{4144sp}%
  \begingroup\makeatletter\ifx\SetFigFont\undefined%
  \gdef\SetFigFont#1#2#3#4#5{%
    \reset@font\fontsize{#1}{#2pt}%
    \fontfamily{#3}\fontseries{#4}\fontshape{#5}%
    \selectfont}%
  \fi\endgroup%
  \begin{picture}(1974,1975)(9198,-1372)
    \put(9724,-1243){\makebox(0,0)[b]{\smash{{\SetFigFont{12}{14.4}{\rmdefault}{\mddefault}{\updefault}{\color[rgb]{0,0,0}\pc{$$[G^-, \varphi_{a}]$$}}%
          }}}}
    \put(9198,254){\makebox(0,0)[lb]{\smash{{\SetFigFont{12}{14.4}{\rmdefault}{\mddefault}{\updefault}{\color[rgb]{0,0,0}\pc{$$\langle \bar j | = \bar\phi_{\bar j}$$}}%
          }}}}
  \end{picture}%
  \caption{The near-boundary region of the shrinking boundary degeneration for $y^a$ derivative, with insertion (\ref{eq:remaining_insertions}) near the boundary. This amplitude vanishes, as described in the text.}
  \label{fig:y_zero_disk}
\end{minipage}
\hspace{0.07\textwidth}
\begin{minipage}[t]{0.5\textwidth}
  \centering
  \begin{picture}(0,0)%
    \includegraphics{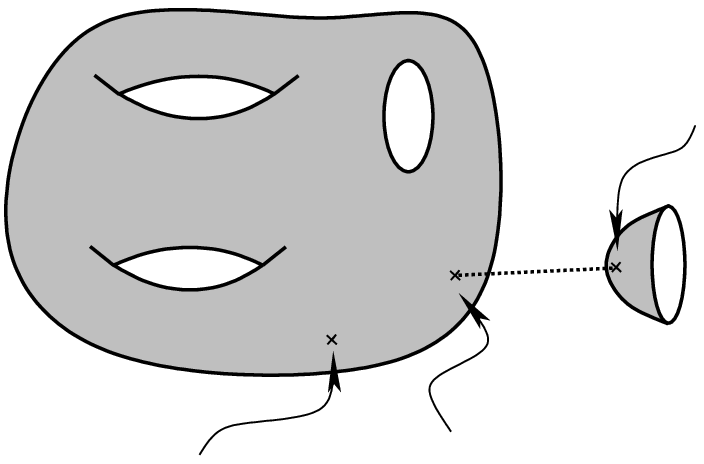}%
  \end{picture}%
  \setlength{\unitlength}{4144sp}%
  \begingroup\makeatletter\ifx\SetFigFont\undefined%
  \gdef\SetFigFont#1#2#3#4#5{%
    \reset@font\fontsize{#1}{#2pt}%
    \fontfamily{#3}\fontseries{#4}\fontshape{#5}%
    \selectfont}%
  \fi\endgroup%
  \begin{picture}(3190,2323)(7836,-5295)
    \put(9983,-4965){\makebox(0,0)[lb]{\smash{{\SetFigFont{12}{14.4}{\rmdefault}{\mddefault}{\updefault}{\color[rgb]{0,0,0}\parbox{0pt}{$$\{G^-, [\overline G^-, \omega_a]\}$$}}%
          }}}}
    \put(8336,-5166){\makebox(0,0)[b]{\smash{{\SetFigFont{12}{14.4}{\rmdefault}{\mddefault}{\updefault}{\color[rgb]{0,0,0}\pc{$$[G^-, \varphi_{a}]$$}}%
          }}}}
    \put(10933,-3415){\makebox(0,0)[lb]{\smash{{\SetFigFont{12}{14.4}{\rmdefault}{\mddefault}{\updefault}{\color[rgb]{0,0,0}\pc{$\bar\omega_{\bar b}$}}%
          }}}}
    \put(10308,-4140){\makebox(0,0)[b]{\smash{{\SetFigFont{12}{14.4}{\rmdefault}{\mddefault}{\updefault}{\color[rgb]{0,0,0}\pc{$g^{a \bar b}$}}%
          }}}}
  \end{picture}%
  \caption{Amplitude for the shrinking boundary degeneration for $y^a$ derivative, with insertion (\ref{eq:remaining_insertions}) elsewhere on the Riemann surface. This is non-zero unless the disk one-point function vanishes.}
  \label{fig:y_shrink_bndry}
\end{minipage}
\end{figure}

Secondly, (\ref{eq:remaining_insertions}) may be near the shrinking boundary. For the $\bar t^{\bar i}$ derivative, the near-boundary region is a disk two-point function in the anti-topological twisting, denoted $\Delta_{\bar i \bar j}$. This term was identified in \cite{Walcher:2007tp}, and is the last term in (\ref{eq:tbar_anomaly}). For the $y^a$ derivative, the near-boundary region is shown in Figure \ref{fig:y_zero_disk}. We can replace the tube with a complete set of closed string ground states, $\sum_{I,\bar J} | I \rangle g^{I\bar J}\langle\bar J |$, where $g^{I\bar J}$ is the $tt^*$ metric, and $I$, $\bar J$ run over all $(c,c)$ and $(a,a)$ chiral primary states, respectively. Standard considerations on the rest of the Riemann surface force $|I\rangle = | i \rangle$ to be a charge $(1,1)$ ({\em i.e.} marginal) state. $\langle \bar J | = \langle\bar j|$ is thus a charge $(-1,-1)$ state from the $(a,a)$ chiral ring. Near the boundary the theory is anti-topologically twisted, making $G^-$ and $\overline G^-$ dimension $1$ as supercurrents, and so allowing contour deformation. Using the properties of the chiral rings, (\ref{eq:remaining_insertions}) can be written as $ \int_\Sigma[G^- + \overline G^-, \varphi_a]$. The contour of $(G^-+\overline G^-)$ can be deformed off the disk, annihilating both $\langle \bar j|$ and the boundary, so this case is zero.

\begin{figure}
  \centering
\begin{picture}(0,0)%
\includegraphics{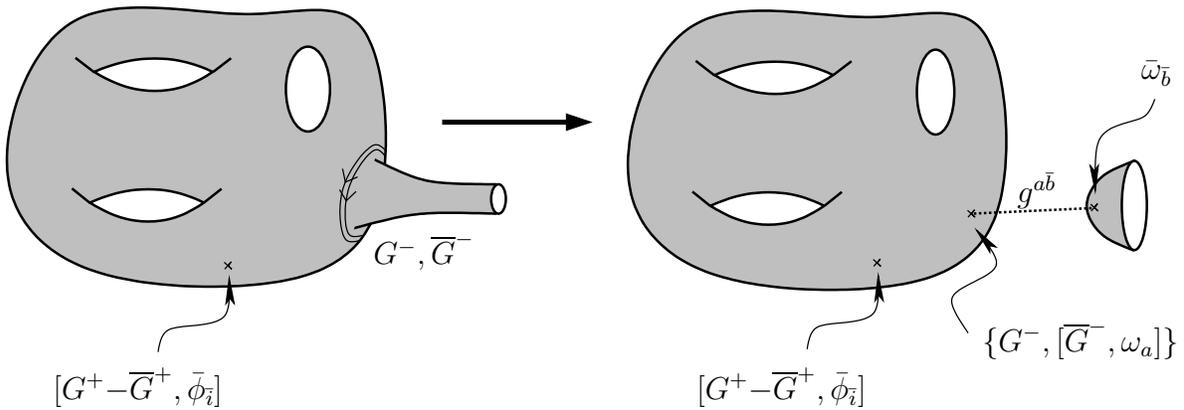}%
\end{picture}%
\setlength{\unitlength}{4144sp}%
\begingroup\makeatletter\ifx\SetFigFont\undefined%
\gdef\SetFigFont#1#2#3#4#5{%
  \reset@font\fontsize{#1}{#2pt}%
  \fontfamily{#3}\fontseries{#4}\fontshape{#5}%
  \selectfont}%
\fi\endgroup%
\begin{picture}(6901,2576)(4125,-5529)
\put(10933,-3415){\makebox(0,0)[lb]{\smash{{\SetFigFont{12}{14.4}{\rmdefault}{\mddefault}{\updefault}{\color[rgb]{0,0,0}\pc{$\bar\omega_{\bar b}$}}%
}}}}
\put(10308,-4140){\makebox(0,0)[b]{\smash{{\SetFigFont{12}{14.4}{\rmdefault}{\mddefault}{\updefault}{\color[rgb]{0,0,0}$g^{a \bar b}$}%
}}}}
\put(9983,-4965){\makebox(0,0)[lb]{\smash{{\SetFigFont{12}{14.4}{\rmdefault}{\mddefault}{\updefault}{\color[rgb]{0,0,0}\parbox{0pt}{$$\{ G^-, [\overline G^-, \omega_a]\}$$}}%
}}}}
\put(6342,-4512){\makebox(0,0)[lb]{\smash{{\SetFigFont{12}{14.4}{\rmdefault}{\mddefault}{\updefault}{\color[rgb]{0,0,0}$G^-, \overline G^-$}%
}}}}
\put(4423,-5250){\makebox(0,0)[b]{\smash{{\SetFigFont{12}{14.4}{\rmdefault}{\mddefault}{\updefault}{\color[rgb]{0,0,0}\pc{$$[G^+ - \overline G^+, \bar\phi_{\bar i}]$$}}%
}}}}
\put(8272,-5250){\makebox(0,0)[b]{\smash{{\SetFigFont{12}{14.4}{\rmdefault}{\mddefault}{\updefault}{\color[rgb]{0,0,0}\pc{$$[G^+ - \overline G^+, \bar\phi_{\bar i}]$$}}%
}}}}
\end{picture}%
  \caption{Amplitude for the shrinking boundary degenerating for $\bar t^{\bar i}$ derivative, with the insertion (\ref{eq:remaining_insertions}) located away from the shrinking boundary. On the right we have replaced the tube with a sum over states $\omega_a$ of charge $(1,2)$ and $(2,1)$, rendering the near-boundary region a disk one-point function.}
  \label{fig:tbar_shrink_bndry}
\end{figure}

Lastly, (\ref{eq:remaining_insertions}) may be inserted somewhere else on the Riemann surface, as shown in Figures \ref{fig:y_shrink_bndry} and \ref{fig:tbar_shrink_bndry} for the $y^a$ and $\bar t^{\bar i}$ derivatives, respectively. The tube is again replaced with a complete set of ground states $\sum_{I,\bar J} | I\rangle g^{I\bar J} \langle \bar J|$. To avoid annihilation by $G^-$ and $\overline G^-$ localised to the tube attachment point, $| I \rangle$ must be in the $(c,c)$ chiral ring and have $q_I, \bar q_I\neq 0$. Furthermore, both of the insertions (\ref{eq:remaining_insertions}) are (linear combinations of) states with $(0,-1)$ or $(-1,0)$ left- and right-moving $U(1)_R$ charge, and the tube end-point moduli contribute charge $(-1,-1)$, so $| I \rangle$ is required to be a (linear combination of) charge $(1,2)$ or $(2,1)$ state. We denote these $\omega_a$, index $a$ running over charge $(1,2)$ and $(2,1)$ chiral primaries. Note that the $\omega_a$ are not associated with marginal deformations of the topological string in question; they are associated with deformations of the mirror. In the A-model, they correspond to target space 3-forms, and hence to complex structure deformation, and in the B-model they are $(1,1)$ forms, and so correspond to K\"ahler deformations. Near the shrinking boundary the resulting amplitude is the disk one-point function,
\begin{equation}
 \overline C_{\bar a} = \langle \bar \omega_{\bar a} | B \rangle.
\end{equation}
This case thus contributes the following new terms to (\ref{eq:tbar_anomaly}) and (\ref{eq:y_anomaly}): for the derivative with respect to $\bar t^{\bar i}$,
\begin{equation}
  \label{eq:tbar_new_term}
  g^{\bar a b}\overline C_{\bar a}\int_{{\cal M}_{g, h-1}}[dm]\left\langle \{G^-, [\overline G^-, \omega_b]\}[G^+-\overline G^+, \bar\phi_{\bar i}]\right\rangle_{\Sigma_{g, h-1}},
\end{equation}
and for the derivative with respect to $y^a$,
\begin{equation}
  \label{eq:y_new_term}
 g^{\bar a b} \overline C_{\bar a}\int_{{\cal M}_{g, h-1}}[dm]\left\langle \{G^-, [\overline G^-, \omega_b]\}[G^-, \varphi_a]\right\rangle_{\Sigma_{g, h-1}},
\end{equation}
where the $m$'s are the moduli of the Riemann surface $\Sigma_{g, h-1}$ --- the corresponding insertions of $G^-$ and $\overline G^-$ folded with Beltrami differentials have been suppressed. Note that the $G^-$ and $\overline G^-$ contours around $\omega_b$ and $\varphi_{ a}$ cannot be deformed as they are dimension $2$ as supercurrents; and that the $(G^+ - \overline G^+)$ contour around $\bar \phi_{\bar i}$ cannot be deformed as it does not annihilate any additional boundaries that may be present.

\section{Discussion}

Our analysis has uncovered additional contributions to both the anti-holomorphic and wrong model dependence of the open topological string partition function, if $\overline C_{\bar a}$ is not zero. The former introduces an additional term (\ref{eq:tbar_new_term}) to (\ref{eq:tbar_anomaly}). This term does not correspond to any of the standard open topological string amplitudes, as the insertion $\bar\omega_{\bar a}$ is not marginal. It is therefore not possible to include this additional term in a simple recursion relation, so much of the power of the standard holomorphic anomaly equation is lost. The wrong model derivative anomaly also has no recursive interpretation, but more crucially it manifestly breaks the decoupling of A- and B-model moduli.

The derivation above may not seem to distinguish between compact and non-compact Calabi-Yau target spaces. In fact, the anomalies need only appear in the compact case, as we demonstrate by example in the next paragraph. Beforehand, note that this agrees with our expectations: D-branes wrapped on cycles in compact Calabi-Yau manifolds and filling spacetime (or perhaps even two directions in spacetime \cite{Walcher:2007qp}) give an inconsistent setup unless there are sinks for the topological D-brane charges. Simultaneously, these sinks cancel the disk one-point functions, and so the appearance of the new anomalies is correlated with invalid spacetime constructions.

Furthermore, the standard results of the Chern-Simons gauge theory and matrix models as open topological field theories are not affected by the new anomalies. Consider, for example, $N$ D-branes wrapping the $S^3$ of the space $T^* S^3$, again giving $\overline C_{\bar a} \neq 0$. The total space of $T^* S^3$ is Calabi-Yau and non-compact, with the $S^3$ radius as the complex structure modulus. It is well-known that open topological string theory on this space is the $U(N)$ Chern-Simons theory, which is topological and should be independent of the $S^3$ radius. To resolve this apparent contradiction, consider embedding $T^* S^3$ in a compact space containing a second 3-cycle in the same homology class as the base $S^3$, wrapped by $N$ anti-D-branes. The boundary states of the two stacks combine to give $\overline C_{\bar a}=0$, and the new anomalies do not appear. Now take the limit where the second 3-cycle moves infinitely far away from the base $S^3$ to recover an anomaly-free local Calabi-Yau construction. The point is that in non-compact Calabi-Yau manifolds, the new anomalies can be removed by an appropriate choice of boundary conditions at infinity.

\begin{acknowledgments}
We thank M.~Aganagic, E.~Getzler, T.~Graber, K.~Hori,
C.~Vafa, K.~Vyas and J.~Walcher for useful discussions. 
This research is supported in part by DOE grant DE-FG03-92-ER40701.
H.O. is also supported in part by the Kavli Foundation and 
by the World Premier International Research Center Initiative (WPI Initiative), MEXT, Japan. 
 
\end{acknowledgments}

\end{document}